\documentclass[12pt]{article}
\usepackage{graphicx}
\usepackage{dcolumn}
\usepackage{bm}
\usepackage[usenames,dvipsnames]{color}
\usepackage{amsmath,amsfonts,amssymb}
\usepackage{fullpage}

\begin{document}

\title{Emergent states in heavy electron materials}

\author{Yi-feng Yang\\
{\small Beijing National Laboratory for Condensed Matter Physics and} \\ 
{\small Institute of Physics, Chinese Academy of Sciences, Beijing 100190, China} \\
{\small E-mail: yifeng@iphy.ac.cn} \\ \\
David Pines\\
{\small Department of Physics, University of California, Davis, California 95616, USA}\\
{\small E-mail: david.pines@gmail.com}
}

\maketitle

\begin{abstract} We obtain the conditions necessary for the emergence of various low temperature ordered states (local moment antiferromagnetism, unconventional superconductivity, quantum criticality, and Landau Fermi liquid behavior) in Kondo lattice materials by extending the two-fluid phenomenological theory of heavy electron behavior \cite{Nakatsuji2004,Curro2004,Yang2008a,Yang2008b,Yang2009,Yang2011,Warren2011,Shirer2012} to incorporate the concept of hybridization effectiveness. We use this expanded framework to present a new phase digram and consistent physical explanation and quantitative description of measured emergent behaviors such as the pressure variation of the onset of local moment antiferromagnetic ordering at $T_N$, the magnitude of the ordered moment, the growth of superconductivity within that ordered state, the location of a quantum critical point, and of a delocalization line in the pressure/temperature phase diagram at which local moments have disappeared and the heavy electron Fermi surface has grown to its maximum size. We apply our model to CeRhIn$_5$ and a number of other heavy electron materials and find good agreement with experiment.  \end{abstract}

Heavy electron materials provide a unique $f$-electron laboratory for the study of the interplay between localized and itinerant behavior. At comparatively high temperatures, itinerancy emerges as the localized $f$-electrons collectively reduce their entropy by hybridizing with the conduction electrons to form a new state of matter, an itinerant heavy electron Kondo liquid that displays scaling  (non-Landau Fermi liquid) behavior. The emergent Kondo liquid coexists with the hybridized spin liquid that describes the lattice of local moments whose magnitude has been reduced by hybridization until one reaches the comparatively low temperatures at which unconventional superconductivity and hybridized local moment antiferromagnetism compete to determine the ground state of the system over a wide regime of pressures. In practice the phases often coexist, as experiments on the 115 (CeRhIn$_5$ \cite{Mito2003,Kawasaki2003,Park2006} and CeCoIn$_5$ \cite{Kenzelmann2008}) and 127 (CePt$_2$In$_7$ \cite{Bauer2010}) Kondo lattice materials demonstrate. A major challenge for the heavy electron community has been finding a consistent framework and simple phenomenological description of their measured emergent behaviors that include the pressure variation of the onset of antiferromagnetic ordering at $T_N$, the growth of superconductivity within that ordered state, the onset of quantum critical behavior, and the growth of the heavy electron Fermi surface. 

While progress has been made on developing a phenomenological theory of Kondo lattice materials \cite{Yang2011}, we do not yet possess a microscopic theory of the emergence of the Kondol liquid as a new quantum state of matter that displays universal behavior below a characteristic temperature\cite{Nakatsuji2004,Curro2004,Yang2008a,Yang2009}, $T^*$,  or a satisfactory general framework for characterizing the conditions necessary for the emergence of the competing low temperature ordered states, since experiments \cite {Yang2008b} have shown that the seminal phase diagram proposed by Doniach\cite{Doniach1977} does not apply to most materials. In this paper we introduce the concept of {\em hybridization effectiveness} as the organizing principle responsible for the emergence of low temperature order and show it leads to a new phase diagram that provides a realistic explanation for the emergence of different low temperature phases for Kondo lattice materials. The role of hybridization effectiveness in reducing the entropy of the lattice of local moments present at high temperatures can be described quantitatively through the pressure dependence of a simple parameter present in the phenomenological two-fluid description \cite{Nakatsuji2004,Curro2004,Yang2008a,Yang2008b,Yang2009,Yang2011,Warren2011,Shirer2012} of heavy electron materials whose magnitude can be determined from fits to experimental data. We show that our new approach makes possible a consistent physical explanation and quantitative description of the low temperature behaviors of a number of heavy electron materials and predicts the presence of a delocalization line in the pressure/temperature phase diagram. It offers promise of serving as a standard model of heavy electron emergent behaviors and providing the basis for the development of a microscopic theory that explains these.

\section{A model for hybridization effectiveness}
In the two-fluid description, the heavy electron Kondo liquid emerges below a characteristic temperature, $T^*$, as a collective hybridization-induced instability of the spin liquid that describes the lattice of local moments coupled to background conduction electrons. $T^*$ is determined by the effective RKKY interaction between the nearest neighbor local moments \cite{Yang2008b}; the emergent scaling behavior of the Kondo liquid is characterized by the heavy electron order parameter \cite{Yang2008a}, 
\begin{equation}
f_h\left(T\right)=f_0\left(1-\frac{T}{T^*}\right)^{3/2},
\label{Eq:fh}
\end{equation}
that extends from $T^*$  to a comparatively low temperature, $T_0$, at which ordering begins to set in \cite{Warren2011}. Over this temperature range it coexists with the "hybridized" quantum spin liquid, whose individual components (that still interact through the inter-site RKKY coupling) possess a spectral weight that diminishes with lowering temperature as $f_l(T)=1-f_h(T)$.

\begin{figure}[t]
\centerline{{\includegraphics[width=.7\textwidth]{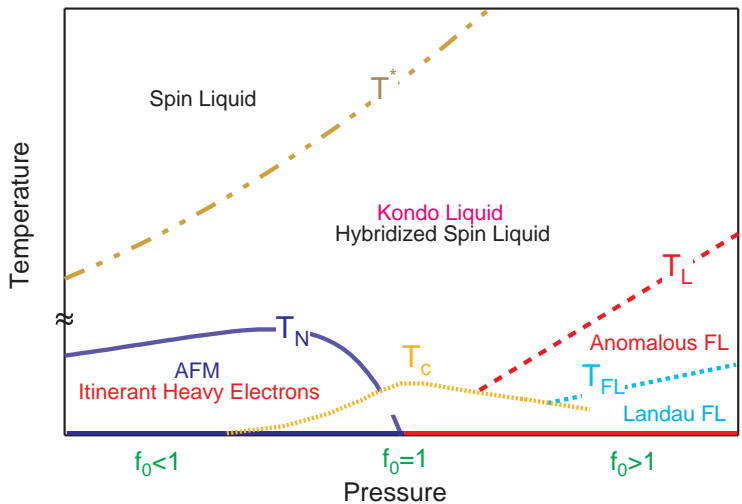}}}
\caption{
{A proposed phase diagram in which a quantum critical point at $f_0=1$ separates weak and strong hybridizing materials: if $f_0<1$, antiferromagnetic local moment order in the presence of heavy electrons sets in below $T_N$; for $f_0>1$, one has complete delocalization of $f$-electrons along a line $T_L$, with the heavy electrons forming a Landau Fermi liquid below $T_{FL}$. Around the quantum critical point at $f_0=1$, the heavy electrons may condense into superconductivity below $T_c$ due to magnetic quantum fluctuations. $T^*$ marks the onset of Kondo liquid emergence produced by collective hybridization.}
\label{Fig:PhaseDiagram}}
\end{figure}

The pressure dependent parameter, $f_0$, in Eq. (\ref{Eq:fh}) provides a direct measure of the ultimate effectiveness of the underlying collective hybridization in reducing the magnitude of strength of each localized $f$-moment. Importantly, as may be seen in the schematic summary of our findings in Fig. \ref{Fig:PhaseDiagram}, it constrains the nature of the ground state. As long as  $f_0<1$, hybridization is always incomplete, so antiferromagnetic local moment order becomes possible at a temperature reduced by hybridization: it takes place in the presence of itinerant heavy electrons that can become superconducting, so that over a wide range of pressures local moment order coexists with superconductivity of magnetic origin. At $f_0=1$, the collective hybridization of local moments becomes complete at T=0; hence if superconductivity is suppressed by application of a strong enough magnetic field, one expects to find a quantum critical point, denoting a $T=0$ transition between $f$-electron local order and a fully itinerant heavy electron state. For $f_0>1$, the hybridization of local moments is complete and the Fermi surface will have grown to its maximum size at some finite temperature, $T_L$; Kondo liquid scaling behavior is expected between $T^*$ and $T_L$, while $T_L$ is fixed by $f_0$ and $T^*$. From Eq. (\ref{Eq:fh}), we have
\begin{equation}
\frac{T_L}{T^*} = 1- f_0^{-2/3}.
\label{Eq:TL}
\end{equation}
For $f_0\sim1$, because of the proximity of a quantum critical point, immediately below $T_L$ one does not expect to find Landau Fermi liquid (FL) behavior; instead, the properties of the heavy electrons will be determined by their coupling to quantum critical fluctuations; anomalous behavior, including further mass enhancement, continues until one reaches the still lower characteristic temperature, $T_{FL}$, shown in Fig. \ref{Fig:PhaseDiagram}, at which transport and other properties are no longer dominated by the coupling of quasi-particles to quantum critical fluctuations and one finds Landau Fermi liquid behavior.

\subsection{The hybridized quantum spin liquid}
The magnetic properties of the hybridized spin liquid can be studied by assuming its dynamic spin susceptibility takes a mean-field form
\begin{equation}
\chi_l(\mathbf{q},\omega)=\frac{f_l\chi_0}{1-zJ_\mathbf{q}f_l\chi_0-i\omega/\gamma_l},
\label{Eq:chi}
\end{equation}
where $\chi_0$ is the local susceptibility of an individual $f$-moment, $z$ is the coordination number, $\gamma_l$ is the local relaxation rate, and $J_\mathbf{q}$ is the RKKY exchange coupling. For simplicity, the magnetic moment of the $f$-spins and the Boltzmann constant are set to unity. The hybridized spin liquid will begin to order at a N\'eel temperature that is, according to Eq. (\ref{Eq:chi}), determined by
\begin{equation}
zJ_\mathbf{Q}f_l(T_N)\chi_0(T_N)=1.
\end{equation}
All local physics can be included in the local moment static susceptibility $\chi_0$. For example, crystal field effects become important when the crystal field splitting from the ground state doublet is smaller than or comparable to $T^\star$; if the crystal field configuration is known and plays a role, it can be easily taken into account in $\chi_0$. In this paper, we take for simplicity $\chi_0=C/T$, where $C$ is the Curie constant. For $T>T^*$, Eq. (\ref{Eq:chi}) reduces to the Curie-Weiss form of the static susceptibility $\chi_l(0, 0)=C/(T+\theta)$, where $\theta=CzJ_{\mathbf{q}=0}$. We have then
\begin{equation}
\frac{T_N}{T^*}=\eta f_l(T_N),
\label{Eq:TN}
\end{equation}
where the parameter $\eta=CzJ_\mathbf{Q}/T^*$ reflects the effect of the crystal lattice and magnetic frustration on $T_N$, while $f_l(T_N)$ reflects the role played by collective hybridization in reducing the N\'eel temperature. $\eta$ is pressure-independent because of the RKKY nature of both $T^*$ and $J_\mathbf{Q}$ \cite{Yang2008b}.

The magnitude of the ordered moment $\mu^2$  is determined by the local moment order parameter:
\begin{equation}
\frac{\mu^2}{\mu_0^2} = f_l(T_N),
\label{Eq:M0}
\end{equation}
where $\mu_0$ is the local moment strength above $T^*$. Where accurate measurements of the strength of the ordered moment in the antiferromagnetic state exist, a good test of our model is to combine Eqs. [\ref{Eq:TN}] and [\ref{Eq:M0}] to obtain an expression that relates three experimentally measurable quantities at a fixed value of $\eta$,
\begin{equation}
T_N= \eta T^*\left(\frac{\mu^2}{\mu_0^2}\right).
\label{Eq:TNM}
\end{equation}
Since $T^*$ increases with increasing pressure, while $\mu$ decreases, it follows that in general the $T_N$ versus pressure curve should exhibit a maximum, as shown in Fig. \ref{Fig:PhaseDiagram}. 

In applying Eqs. [\ref{Eq:TN}] and [\ref{Eq:M0}] to experiment, one needs to take relocalization effects into account. As discussed in apRoberts-Warren et al. \cite{Warren2011} and Shirer et al. \cite{Shirer2012}, Knight shift experiments demonstrate that Kondo liquid scaling ends at a temperature $T_0$, below which local moment ordering begins to win out over collective hybridization in determining the temperature evolution of $f_l(T)$. As the temperature is lowered, the rate of collective hybridization is first reduced and then reversed, i.e., the spectral weight of $f_l(T)$ begins to increase as the temperature is further lowered. For the two materials of which this relocalization has been explored in detail \cite{Warren2011,Shirer2012}, it turns out that $f_l(T_N)\sim f_l(T_0)$ and $T_0\sim2T_N$.

\begin{figure}[t]
\centerline{{\includegraphics[width=.8\textwidth]{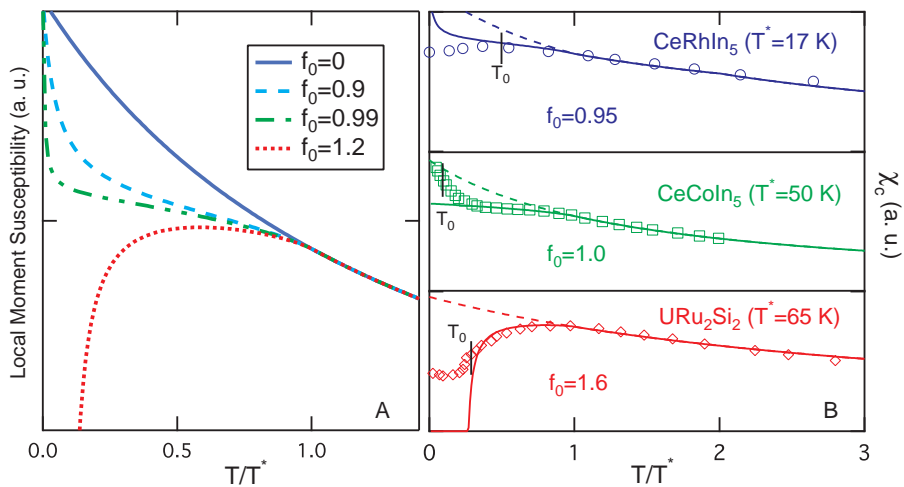}}}
\caption{
{The role played by hybridization effectiveness in determining the static magnetic susceptibility. (A) Schematic illustration of the influence of $f_0$ on $\chi_l(0,0)$ for a fixed value of $\theta/T^*=1.4$. (B) A comparison of our calculated and experimental values of the c-axis susceptibility $\chi_c$ for three materials: CeRhIn$_5$ \cite{Hegger2000}, CeCoIn$_5$ \cite{Curro2001} and URu$_2$Si$_2$ \cite{Yokoyama2002}. The solid lines are theoretical fits with $\theta/T^*=1.4$, 1.35, 3.5, respectively; also shown there is the temperature, $T_0$, at which scaling behavior is altered by the onset of low temperature order.}
\label{Fig:chi}}
\end{figure}

\subsection{Determining $f_0$} We can use Eq.~(\ref{Eq:chi}) to determine $f_0$ from measurements of the static spin susceptibility if we make the physically reasonable assumption that the hybridized spin liquid contribution is dominant for a broad range of temperatures below $T^*$. Such approximation is reasonable due to the large magnetic response of the local moments seen at temperature above $T^*$. Fig.~\ref{Fig:chi} shows the evolution of $\chi_l(0,0)$ with $f_0$ and the fit to experimental data in several compounds -- from the antiferromagnet CeRhIn$_5$ \cite{Hegger2000} to the superconductor CeCoIn$_5$ \cite{Curro2001} to the 5f compound URu$_2$Si$_2$ that exhibits hidden order \cite{Yokoyama2002}. We find $f_0=0.95$ for CeRhIn$_5$, 1.0 for CeCoIn$_5$ and 1.6 for URu$_2$Si$_2$. With increasing $f_0$, we find a crossover from peak to plateau to peak structure in the uniform susceptibility of exactly the kind that has been observed in many heavy electron materials. However, the origin of the peak seen in materials with $f_0<1$, such as CeRhIn$_5$, and those in which $f_0>1$, such as URu$_2$Si$_2$, is different. For $f_0>1$, the peak comes from rapid delocalization of the $f$-moments at $\sim T^*$; for $f_0<1$, it is related to the onset of relocalization of the Kondo liquid as a precursor of the antiferromagnetic ordering at $T_0\ll T^*$ \cite{Warren2011,Shirer2012}.

\subsection{The Kondo liquid}
The Kondo liquid contributes to the magnetic entropy and other material properties that display anomalous behavior, as well as physical quantities such as transport and superconductivity that originate uniquely in the itinerant $f$-electrons. The Kondo liquid entropy $S_h$ can be obtained by integrating over the universal scaling expression for its specific heat $C/T\sim[1+\ln(T^*/T)]$ \cite{Yang2008a} and recognizing that continuity at $T^*$ requires that $S_h(T^*)=S_l(T^*)=R\ln2$ ($R$ is the gas constant). When the scaling behavior of the Kondo liquid ends at a temperature $T_x$ ($=T_L$ if $f_0>1$), the two-fluid model predicts that the heavy electron entropy will be given by \cite{Yang2011}
\begin{equation}
S_{h}(T_x)=R\ln 2\frac{T_x}{2T^*}\left[2+\ln\left(\frac{T^*}{T_x}\right)\right].
\end{equation}
Below $T_L$, no local moments are present; only heavy electrons are responsible for the measured system behavior and a one-fluid picture is recovered, as may be verified in Knight shift measurement. Their Fermi liquid specific heat coefficient, $\gamma_h$, at lower temperatures can then be estimated
\begin{equation}
\gamma_{h}\sim\frac{S_h(T_L)}{T_L}=\frac{R\ln 2}{2T^*}\left[2-\ln\left(1-f_0^{-2/3}\right)\right].
\label{Eq:gamma}
\end{equation}
In general, even when $f_0<1$ or $T>T_L$, the specific heat coefficient can still be calculated using Eq. (\ref{Eq:S}).

\section{Application to CeRhIn$_5$ }

\begin{figure}[t]
\centerline{{\includegraphics[width=.7\textwidth]{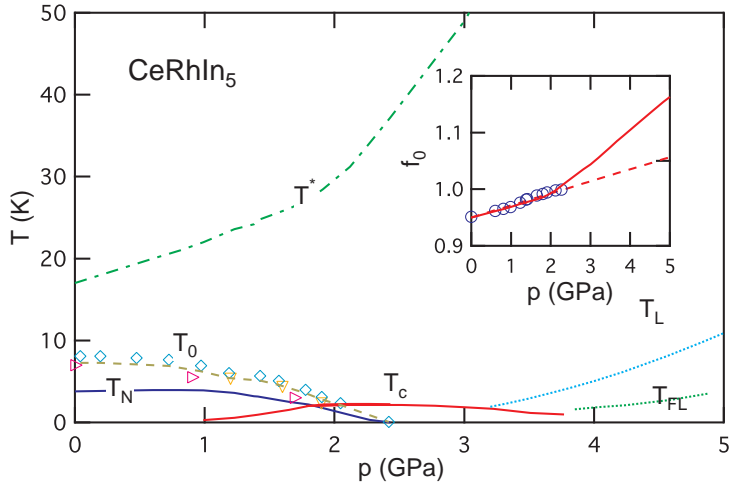}}}
\caption{
{The proposed phase diagram of CeRhIn$_5$. The pressure dependence of $T_N$ and $T_c$ are taken from \cite{Park2008,Park2009}, while the coherence temperature $T^*$ is that determined by Yang {\em et al.} \cite{Yang2008a,Yang2008b,Yang2011}. The experimental points determining $T_0$, the cut-off temperature for Kondo liquid scaling, reflect its signature in several different experimental probes: a pseudo-gap like feature in the spin-lattice relaxation rate \cite{Kawasaki2002}, a change in the anomalous Hall effect \cite{Nakajima2007} and inelastic neutron scattering spectrum \cite{Bao2002}, and a peak in the measured ''anomalous'' Kondo liquid susceptibility \cite{Hegger2000}. $T_L$ is the delocalization temperature expected if $f_0(p)\sim J\rho$, where $J$ is the local Kondo coupling and $\rho$ is the density of states of the conduction electrons. The inset shows the hybridization parameter $f_0(p)$ as a function of pressure: the points are derived from $T_N(p)$; the solid line is its behavior at higher pressures if $f_0$ scales with $J\rho$ in the vicinity of the quantum critical point; the dashed line is a simple linear extrapolation.}
\label{Fig:CeRhIn5}}
\end{figure}

CeRhIn$_5$ provides an excellent proving ground for the approach we have described.
Our proposed phase diagram for CeRhIn$_5$ is given in Fig.~\ref{Fig:CeRhIn5}, and resembles closely the experimental pressure-temperature phase diagram derived from the resistivity exponent \cite{Park2008}. The Knight shift experiments of Shirer et al. \cite{Shirer2012} show that $f_l(T_N)\approx f_l(2T_N)$ at svp; we assume this to be true in the entire pressure range in the following analysis, with $f_l(2T_N)$ being given by the scaling formula in Eq.~(\ref{Eq:fh}).

\subsection{Antiferromagnetism}
\begin{figure}[t]
\centerline{{\includegraphics[width=.7\textwidth]{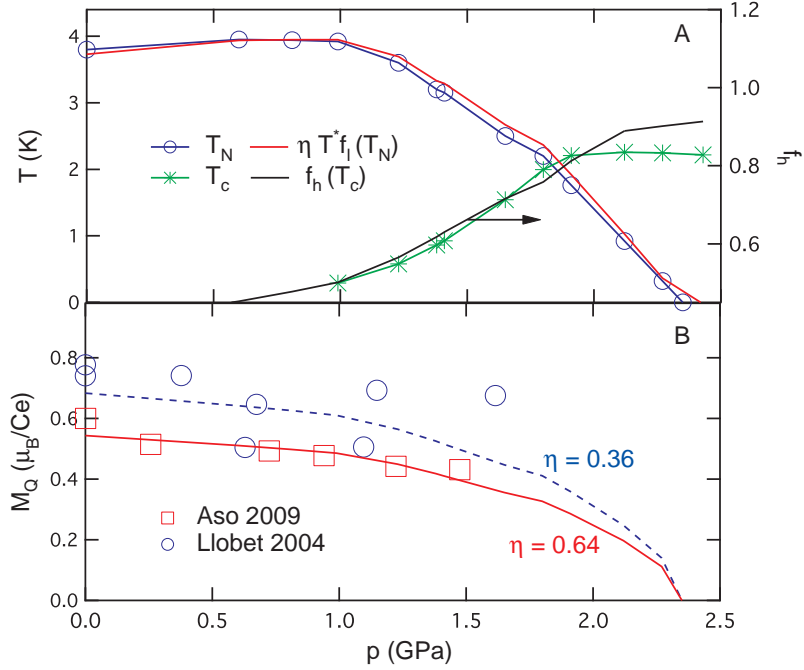}}}
\caption{
{The N\'eel temperature, superconducting transition temperature, and ordered moments as a function of pressure in CeRhIn$_5$. (A) Comparison between experimental \cite{Park2008,Park2009} $T_N(p)$ and $\eta T^* f_l(T_N)$, and between $T_c(p)$ and $f_h(T_c)$ derived from $f_0(p)$ in the inset of Fig.~\ref{Fig:CeRhIn5} for CeRhIn$_5$. With $f_0=0.95$, the best fit requires $\eta=0.64$ in the presence of relocalization. (B) The pressure evolution of the Ce staggered moment at $\textbf{Q}=(\pi,\pi)$. Open squares are data reported by Aso et al. \cite{Aso2009} and open circles are reproduced from Llobet et al. \cite{Llobet2004}. The dashed line is our theoretical fit with $\eta=0.36$ and the solid lines is the fit for Aso's data  with $\eta=0.64$. For Llobet's data, the best fit requires $\eta=0.32$.}
\label{Fig:MQ}}
\end{figure}

On applying Eqs.~[\ref{Eq:TN}], [\ref{Eq:M0}] and [\ref{Eq:TNM}] to CeRhIn$_5$, we obtain the  results for the variation of the N\'eel temperature with pressure, and of the magnitude of the ordered moment found in the N\'eel state that are given in Fig.~\ref{Fig:MQ}. Despite its simplicity, our model yields results in good agreement with experiment \cite{Park2009,Llobet2004,Aso2009}. Fig. \ref{Fig:MQ}A compares the experimental data of $T_N$ and $\eta T^*f_l(2T_N)$. The best fit to the N\'eel temperature yields $\eta=0.36$ for $f_0=0.95$. On taking the strength of the local moments above $T^*$ to be $0.92\mu_B$ \cite{Christianson2002}, $\eta=0.36$ gives the right magnitude of the ordered moment when Eq. (\ref{Eq:TNM}) is used together with the pressure variation of $T_N$. A corresponding best fit to the results of Aso et al. \cite{Aso2009} and Llobet et al. \cite{Llobet2004} requires that $\eta=0.64$ and 0.32, respectively. Our results are seen to be more consistent with those of Llobet et al. \cite{Llobet2004}.

In our mean-field approach the influence of the crystalline lattice and magnetic frustration are incorporated in the coordination number $z$ and $\eta=CzJQ/T^*$. To illustrate a reasonable range for these parameters we consider a 2D square lattice, for which the RKKY coupling has the usual form $J_q=J_n[\cos(q_x)+\cos(q_y)]+2J_{nn}\cos(q_x)\cos(q_y)$, where $J_n$ and $J_{nn}$ are the nearest and next-nearest-neighbor coupling, respectively. We have $\theta=2Cz|J_n+J_{nn}|$ and $\eta T^*=2Cz|J_n-J_{nn}|$ for $Q=(\pi,\pi)$. If we take the values $\eta=0.36$ and $\theta/T^*=1.4$ obtained in the susceptibility fit in Fig.~\ref{Fig:chi}B, we find a frustration ratio $J_{nn}/J_n\approx 0.6$. This indicates a $T^*/2Cz|J_n|\sim 1.1$, consistent with its origin in nearest neighbor RKKY coupling \cite{Yang2008b}.

\subsection{Itinerant electrons and superconductivity within the N\'eel state}
Our model provides a natural explanation for the appearance of superconductivity within the N\'eel state and of the pressure dependence of its transition temperature, $T_c$. Quite generally, local moment ordering takes place before the relocalization process is complete. Below $T_N$, a substantial number of heavy electrons coexist with the ordered local moments in the antiferromagnetic phase, a fraction that increases as the pressure increases. For example in CeRhIn$_5$, despite relocalization, $f_h(T_N)\sim0.37$ at svp. How do these heavy electrons lose their entropy within the N\'eel state? There are two possibilities -- that the relocalization process continues within the N\'eel state, or that the heavy electrons order, most likely in an unconventional superconducting state, since there are ample magnetic fluctuations present that can bring this about \cite{Monthoux2007}. Both likely occur in practice. 

Where superconductivity emerges, it is appealing to argue that to a first approximation $T_c$ scales with the density of heavy electrons present, an argument familiar from the well-known Uemura plot for the cuprate superconductors \cite{Uemura1989}. We test this hypothesis by comparing the corresponding growth in $T_c$ with pressure with the growth in the density of heavy electrons. As may be seen in Fig. \ref{Fig:MQ}A, the increase in $T_c$ seen from 1 GPa to 1.75 GPa mirrors the increase in the density of heavy electrons able to participate in superconductivity, $f_h(T_c)\sim f_h(2T_N)$ assuming no further relocalization below $T_N$. Further confirmation of this view of Kondo liquid behavior comes from the scaling of the Knight shift anomaly above and below $T_c$ seen in CeCoIn$_5$ \cite{Yang2009}, and the observation that $T_c$ is maximum in the vicinity of the critical pressure at which some part of the heavy electron Fermi surface can begin to localize.

\begin{figure}[t]
\centerline{{\includegraphics[width=.7\textwidth]{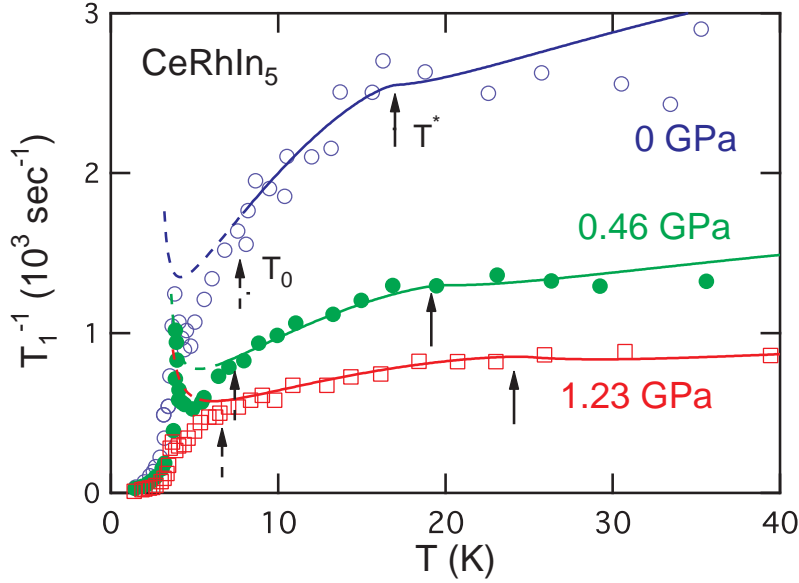}}}
\caption{
{Comparison of measured and calculated NQR spin-lattice relaxation rate at ambient pressure, 0.46$\,$GPa and 1.23$\,$GPa in CeRhIn$_5$ \cite{Kawasaki2002}. The solid lines are theoretical fit with $T^*=17\,$K, 19$\,$K, $24\,$K and $x=0.08$, 0.12, 0.2, respectively.}
\label{Fig:T1}}
\end{figure}

\subsection{Fermi liquid regime} At pressures $p>2.35\,$GPa, the hybridization parameter $f_0>1$, so the ground state of CeRhIn$_5$ should be a Fermi liquid if one applies a magnetic field strong enough to suppress superconductivity. At $T_L=(1-f_0^{-2/3})T^*$, when hybridization is complete, our model predicts a heavy electron specific heat that depends only on $T^*$ and $f_0$ and is given by Eq.~(\ref{Eq:gamma}). $T_L$ can be determined from a measurement of the Fermi surface, which should reach its maximum size there. Our analysis at low pressures suggests that $f_0$ may scale with the Kondo coupling, $J$, in the vicinity of the quantum critical point. If this scaling extends to higher pressures, it leads to the results shown for $T_L$ in Fig.~\ref{Fig:CeRhIn5}. Below $T_L$, the $f$-electrons form a Fermi liquid, but its quasi-particles, being scattered by quantum critical spin fluctuations, do not exhibit the classic properties proposed by Landau until one reaches a lower temperature $T_{FL}$. 

\subsection{Spin-lattice relaxation rate} Nuclear magnetic/quadrupole resonance (NMR/NQR) experiments on the pressure dependence of the spin-lattice relaxation rate provide a further confirmation of the approach developed here. We assume that the spin-lattice relaxation rate between $T^*$ and $T_0$ is dominated by local moment behavior and calculate it using the Moriya formula \cite{Moriya1956}, 
\begin{equation}
\frac{1}{T_1}=\gamma^2T\lim_{\omega\rightarrow 0}F(\mathbf{q})^2\frac{\chi^{''}_l(\mathbf{q},\omega)}{\omega},
\end{equation}
where $\gamma$ is the gyromagnetic ratio and $F(\mathbf{q})^2$ is the form factor. Following \cite{Curro2003}, we use $F(\mathbf{q})^2=F_0^2[\cos^2(\mathbf{q}_x/2)\cos^2(\mathbf{q}_y/2)+x^2\sin^2(\mathbf{q}_x/2)\sin^2(\mathbf{q}_y/2)]$], where $x$, the only free parameter, defines the anisotropy of the hyperfine coupling. On using the values of $f_0(p)$ given in the inset of Fig.~\ref{Fig:CeRhIn5} and assuming that $J_q$ is close to its value for a 2D square lattice, we obtain the fit to the NQR data in CeRhIn$_5$ shown in Fig.~\ref{Fig:T1}. We see there the expected change in slope due to hybridization for $T_1^{-1}$ and that our model provides a fit to experiment above $T_0$, with $T^*=17\,$K at $0\,$GPa, $20\,$K at $0.46\,$GPa and $25\,$K at $1.23\,$GPa, respectively, and an anisotropy parameter $x$ at svp of 0.08 that increases to 0.12 at 0.46 GPa and 0.2 at 1.23 GPa, in agreement with earlier estimates \cite{Curro2003}. The behavior found above $T^*$ reflects the comparatively weak local moment interaction; the temperature dependence seen between $T_0$ and $T_N$ originates from the precursor antiferromagnetic fluctuations \cite{Bao2002} and Kondo liquid relocalization \cite{Warren2011} and cannot be calculated using the present mean-field approximation. The local relaxation rate $\gamma_l$ has been assumed to be a constant in the fit; a temperature dependent $\gamma_l$ may yield improved agreement between theory and experiment. 

\subsection{Magnetic entropy} The magnetic entropy is another important quantity of experimental interest. In many heavy electron antiferromagnets, one finds a small entropy release at $T_N$ despite the presence of a large ordered moment. As will be shown below, our expanded two-fluid model explains these apparently contradictory results as a direct consequence of hybridization and yields agreement with experiment for the magnetic entropy of a number of heavy electron materials.

Above $T^*$, the magnetic entropy is that of localized $f$-electron moments in a ground state doublet that is assumed to have the full magnetic entropy $R\ln2$. At lower temperatures, both the local moments and the Kondo liquid contribute, and the total magnetic entropy is given by \cite{Yang2011}
\begin{equation}
S(T)=R\ln 2\left[f_l(T)+f_h(T)\frac{T}{2T^*}\left(2+\ln\frac{T^*}{T}\right)\right].
\label{Eq:S}
\end{equation}
We expect Eq. (\ref{Eq:S}) to be valid down to a temperature $T_0$, at which low temperature order begins to compete with hybridization in reducing the local moment entropy. Above $T_0$ and for arbitrary $f_0$, where Eq.~(\ref{Eq:gamma}) does not apply, the specific heat coefficient can still be estimated from the temperature derivative of Eq.~(\ref{Eq:S}). For CeRhIn$_5$, as is shown in Fig.~\ref{Fig:CeRhIn5}, a number of different experimental probes suggest that $T_0$ is of order of $2T_N$; importantly the measurement by Shirer et al. of the Knight shift anomaly show that Kondo liquid behavior is strongly suppressed below $T_0$, indicating a rapid relocalization of itinerant heavy electrons \cite{Warren2011,Shirer2012}.

\begin{figure}[t]
\centerline{{\includegraphics[width=.75\textwidth]{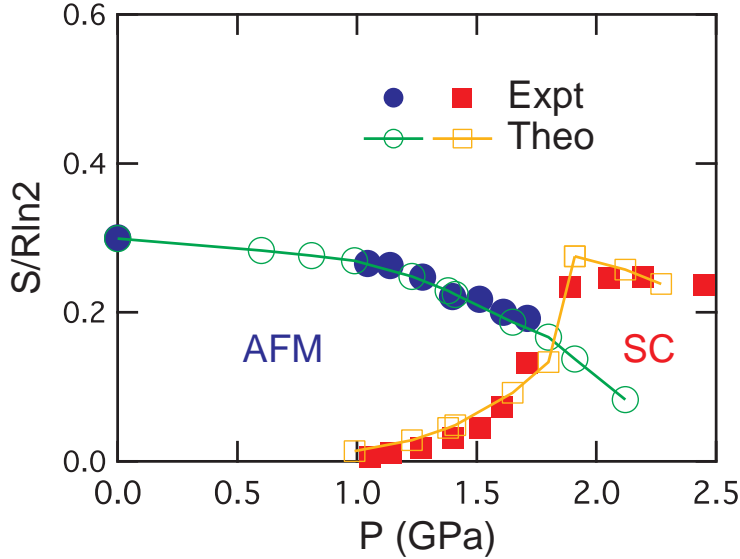}}}
\caption{
{Comparison of measured \cite{Park2009} and calculated magnetic entropies at the N\'eel temperature, $T_N$, and superconducting transition temperature, $T_c$, in CeRhIn$_5$ as a function of pressure \cite{Hegger2000,Park2009}. Theoretical predictions are obtained by using $f_0(p)$ derived in the inset of Fig.~\ref{Fig:CeRhIn5}.}
\label{Fig:S}}
\end{figure}

The physics below $T_0$ requires that Eq.~(\ref{Eq:S}) be modified. In addition to a deviation from scaling in the order parameters $f_h$ and $f_l$, the contribution of each component to the magnetic entropy is expected to be reduced from its value at $T_0$ by the approach of antiferromagnetic order. We find on comparing the magnetic entropy calculated from Eq.~(\ref{Eq:S}) with experiment for CeRhIn$_5$ \cite{Park2009} that 
a factor of $r_N\sim0.35$ has to be introduced at svp in order to obtain agreement with experiment; a reduction of this size is justified by the similar ratio $S(T_N)/S(T_0)\sim 0.4$ found experimentally \cite{Park2009}. As may be seen in Fig.~\ref{Fig:S}, with this factor one obtains good agreement with experiment for pressures up to 1.75 GPa. In contrast, at higher pressures, where superconductivity takes over and the Kondo liquid scaling persists down to $T_c$, no corrections are needed to Eq.~(\ref{Eq:S}), which yields good agreement with experiment for the entropy at $T_c$. 

In calculating the entropy at the superconducting transition deep within the antiferromagnetic phase  ($1.0<p<1.75\,$GPa) shown in Fig.~\ref{Fig:S}, we have excluded the contribution from the ordered moments, whose fraction, $a(T_c)=(1-T_c/T_N)^{0.5}$, has been measured \cite{Llobet2004}, and have assumed the Kondo liquid specific heat does not change between $T_N$ and $T_c$. The entropy release of the coexisting state is then given by,
\begin{equation}
S(T_c)=r_Nf_l(T_N)[1-a(T_c))]R\ln 2 +r_Nf_h(T_N)S_h(T_N)\frac{T_c}{T_N},
\end{equation}
and agrees well with experiment.

At first sight, given the comparatively large magnitude of the ordered moment in CeRhIn$_5$ ($0.6-0.8\mu_B$), the entropy released by antiferromagentic ordering of the local moments appears unexpectedly small, being only $\sim30\%R\ln2$ \cite{Hegger2000}. A little thought shows that the reduction comes in three sequential stages, corresponding to the entropy release at $T_N$, $T_0$ and $T^*$, respectively. When antiferromagnetism is destroyed at $T_N=3.8\,$K, only a fraction $\sim63\%$ of the spectral weight of the strongly hybridized $f$-moments is available to become ''free''; the rest ($\sim37\%$) of that spectral weight has already gone to the Kondo liquid, whose entropy is already much reduced due to strong entanglement between conduction and $f$-electrons. It follows that above $T_N$, the entropy of the ''freed'' moments is then only released at a higher temperature $T_0\sim8\,$K due to strong magnetic correlations close to the phase transition \cite{Shirer2012}, while the total spin entropy will not be fully recovered until all the correlated Kondo liquid has been transformed into uncorrelated $f$-moments above $T^*\sim17\,$K. The continuous and slight change of the magnetic entropy at the ordering temperature as one moves from antiferromagnetism to superconductivity with increasing pressure range requires a gradual change of character of $f$-moments and argues against a dramatic change of hybridization across the quantum critical point.

\section{Other Materials}
\begin{figure}[t]
\centerline{{\includegraphics[width=.6\textwidth]{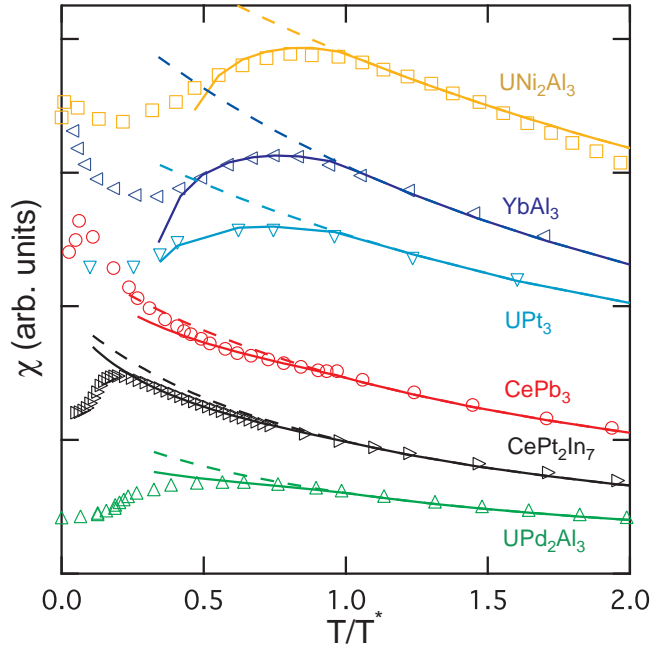}}}
\caption{
{Magnetic susceptibility of several prototypical heavy electron materials. The solid lines are our theoretical fit using the mean-field formula and the dashed lines are the Curie-Weiss susceptibility. The fitting parameters are shown in the text.}
\label{Fig:chi2}}
\end{figure}

\begin{figure}[t]
\centerline{{\includegraphics[width=.7\textwidth]{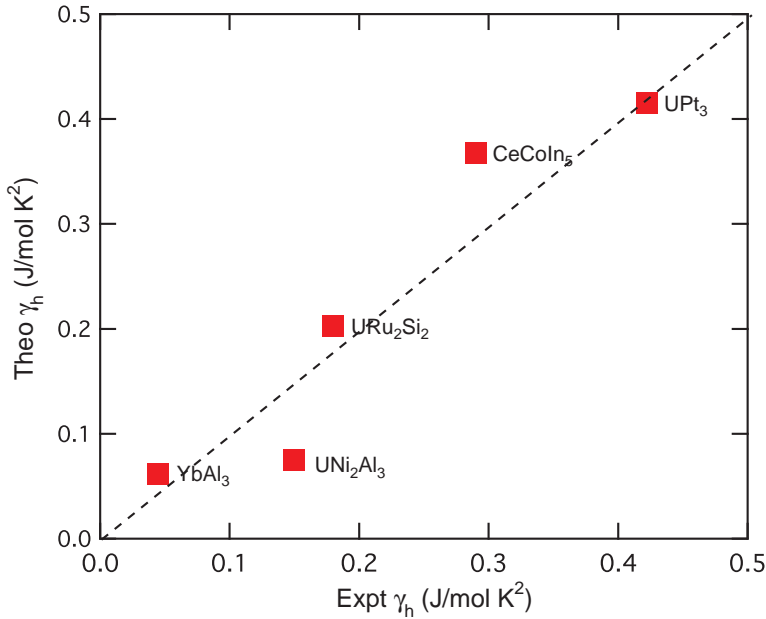}}}
\caption{
{A comparison between the values of the specific heat coefficient predicted by our model and experiment for materials with $f_0\ge1$ \cite{Havinga1973,Visser1986,Geibel1991,Petrovic2001,Yokoyama2002}. }
\label{Fig:gamma}}
\end{figure}

In applying our model to other prototypical heavy electron materials, we focus on the relation between $f_0$ and the magnetic/non-magnetic ground state and the entropy release of the ordered state (if $f_0<1)$ or the residual specific heat coefficient of the Fermi liquid (when $f_0\ge1$), since these only require experimental input at ambient pressure.  Our results are summarized in Fig. \ref{Fig:chi2} and Table \ref{Tab:f0}, and discussed in the subsections below. Because they are strong hybridizers ($f_0>1$), we conclude that UNi$_2$Al$_3$ must be a spin density wave antiferromagnet in which antiferromagnetic order originates from Fermi surface nesting of the heavy electron Kondo liquid \cite{Tateiwa1998}, while the hidden order transition in URu$_2$Si$_2$ must be due to a Kondo liquid instability \cite{Mydosh2011}. As may be seen in Fig.~\ref{Fig:gamma}, our model (Eq.~(\ref{Eq:gamma})) also yields results for the low temperature Fermi liquid specific heat of a number of materials that are surprisingly close to their experimental values \cite{Havinga1973,Visser1986,Geibel1991,Petrovic2001,Yokoyama2002}, which demonstrates that for these materials the average heavy electron effective mass is determined by the onset of collective hybridization at $T^*$ and its effectiveness, $f_0$.

\begin{table}[t]
\centering
\caption{Hybridization effectiveness and low temperature order}
\label{Tab:f0}
\begin{tabular*}{\hsize}{@{\extracolsep{\fill}}ccccc}
 Materials & $T^* (K)$ & $T_{N/L} (K)$ & $f_0$ & Order\\
\hline
\multicolumn{5}{c}{Local moment order}\\
\hline
CePt$_2$In$_7$ & 41 & 5.6 & 0.4 & AFM \\
CePb$_3$ & 16 & 1.1 & 0.5 & AFM \\
UPd$_2$Al$_3$ & 60 & 14.3 & 0.8 & AFM \\
CeRhIn$_5$ & 17 & 3.8 & 0.95 & AFM \\
\hline
\multicolumn{5}{c}{Kondo liquid order}\\
\hline
CeCoIn$_5$ & 50 & 0 & 1.0 & SC \\
UPt$_3$ & 25 & 5 & 1.4 & SC \\
YbAl$_3$ & 160 & 38 & 1.5 & FL\\
URu$_2$Si$_2$ & 65 & 17.5 & 1.6 & HO \\
UNi$_2$Al$_3$ & 120 & 39 & 1.8 & SDW\\
\hline
\end{tabular*}
\end{table}

\subsection{CeCoIn$_5$} For this material, the Kondo liquid temperature $T^*=50\,$K has been determined using a number of experimental probes \cite{Yang2008a,Yang2008b}. On using Eq.~(\ref{Eq:chi}), we show in Fig.~\ref{Fig:chi}B that $f_0\sim1$ provides a good fit to the susceptibility data \cite{Petrovic2001,Curro2001}, a conclusion that is consistent with the expectation that CeCoIn$_5$ is close to a magnetic quantum critical point \cite{Sidorov2002}. We find $S(T_c)=0.18R\ln2$, close to the experimental value of $0.2R\ln2$ at $T_c=2.3\,$K \cite{Petrovic2001}. Taking $T_c$ as the cut-off temperature, the predicted specific heat coefficient at $T_c$ and zero field using Eq.~(\ref{Eq:S}) is $\sim 368\,$mJ/mol K$^2$, comparable to the experimental value of $\sim 290\,$mJ/mol K$^2$ \cite{Petrovic2001}.

\subsection{URu$_2$Si$_2$} 
As may be seen in Fig.~\ref{Fig:chi}B, our best fit to the c-axis susceptibility of URu$_2$Si$_2$ gives the hybridization parameter $f_0=1.6$ and $T_L\approx T_{HO}=17.5\,$K. Below $T_L$, all $f$-electrons become itinerant so that a single component picture is recovered and the Knight shift becomes once again proportional to the susceptibility. This allows us to subtract the spin susceptibility of the two components without uncertainty and the detailed analysis given in \cite{Shirer2012} yields similar results for $f_0$ and $T_L$. These suggest that most local moments have disappeared by the time one reaches $T_{HO}$, so that the hidden order transition must have an itinerant and strongly correlated nature, consistent with experimental observation of gap opening \cite{Bonn1988,Aynajian2010} and Fermi surface reconstruction at $T_{HO}$ \cite{Syro2009}. Our predicted entropy release at $T_{HO}$ is $\sim0.44R\ln2$, somewhat larger than the experimental value of $\sim0.3\,$mJ/mol K$^2$ \cite{Yokoyama2002}. The predicted specific heat coefficient without hidden order is $\sim 203\,$mJ/mol K$^2$, close to the experimental extrapolated value of $\sim 180\,$mJ/mol K$^2$.

\subsection{CePt$_2$In$_7$} Using $\theta=T^*=41\,$K, we find $f_0=0.4$ from a fit to the susceptibility, in which only the local moment contribution is included. In an earlier analysis that included the Kondo liquid contribution and assumed a Wilson ratio $R_W=2$  \cite{Warren2011}, $f_0$ was found to be $\sim1$, a result in obvious contradiction with the magnetic ground state. The contradiction leads us to conclude that one should not in general assume that $R_W=2$, but rather determine it on a case by case basis. For this material, NMR measurements point to a relocalization temperature $T_0\sim14\,$K, below which the magnetic entropy is strongly suppressed. Our scaling result yields $S(T_N)=0.76R\ln2$, which is very much larger than the experimental value of $\sim0.2R\ln2$ \cite{Bauer2010}: just as was the case with CeRhIn$_5$, relocalization effect are large between $T_0=14\,$K and $T_N=5.6\,$K, a result that is consistent with the NMR experimental results \cite{Warren2011}.

\subsection{CePb$_3$} The susceptibility of CePb$_3$ has a plateau below $T^*=16\,$K, followed by a peak before the magnetic transition at $T_N=1.1\,$K \cite{Durkop1985}. Our best fit gives $f_0\sim 0.5$ and $\theta\sim 25\,$K and yields an entropy release of $\sim 0.62R\ln2$ at $T_N$, which is large compared to the experimental value of $0.32 R\ln2$, a discrepancy that is likely due to the rapid suppression of the magnetic entropy between $\sim3\,$K and $T_N$ (see Fig. 8 in \cite{Durkop1985}). 

\subsection{UPt$_3$}
Our fit to the susceptibility of UPt$_3$ gives $f_0=1.4$, $\theta=71\,$K and $T^*=25\,$K. We find $T_L\sim5\,$K, consistent with the deviation from scaling seen in the Knight shift anomaly \cite{Yang2008a}, and $\gamma_h\sim 415\,$mJ/mol K$^2$, in good agreement with the experimental value of $422\,$mJ/mol K$^2$ \cite{Visser1986}.

\subsection{U$M_2$Al$_3$} UPd$_2$Al$_3$ and UNi$_2$Al$_3$ are both antiferromagnetic at low temperatures but, as our fit to the susceptibility shows, its physical origin is different. Because of crystal field effects, for UPd$_2$Al$_3$ \cite{Yang2008b,Grauel1992} we can only give a rough estimate of $\theta=T^*=60\,$K and $f_0=0.8$, while for UNi$_2$Al$_3$ \cite{Geibel1991}, we find $T^*=120\,$K, $\theta=420\,$K and $f_0=1.8$. These quite different values of $f_0$ strongly suggest that the magnetic ground state of UPd$_2$Al$_3$ originates in ordering of its local moments, while that of UNi$_2$Al$_3$ must reflect spin-density-wave ordering that originates in the Fermi surface nesting of the itinerant Kondo liquid.

Since for both materials, the specific heat is enhanced by antiferromagnetic fluctuations only in a narrow range above $T_N$, we can safely neglect the relocalization correction; on doing so we obtain $S(T_N)\sim0.69R\ln2$ for UPd$_2$Al$_3$ and $0.05R\ln2$ for UNi$_2$Al$_3$, in good agreement with the experimental values of $0.67R\ln2$ for UPd$_2$Al$_3$ and $0.12R\ln2$ for UNi$_2$Al$_3$ \cite{Tateiwa1998}. For UNi$_2$Al$_3$, we find $T_L=39\,$K and $\gamma_h=75\,$mJ/mol K$^2$, somewhat smaller than the experimental value of $150\,$mJ/mol K$^2$ \cite{Geibel1991}.

\subsection{YbAl$_3$}
Our fit to the susceptibility of YbAl$_3$ \cite{Havinga1973} gives $f_0=1.5$, $\theta=340\,$K and $T^*=160\,$K. This is consistent with the nonmagnetic ground state of this compound for which we find $T_L\sim38\,$K. The specific heat coefficient is estimated to be $\gamma_h\sim 62\,$mJ/mol K$^2$, consistent with the experimental value of $45\,$mJ/mol K$^2$. 

\subsection*{}
For materials for which $f_0$ cannot be easily obtained, one can still estimate the entropy release by making the approximation $f_0=1$ in Eq. (\ref{Eq:S}). For antiferromagnetic compounds with $f_0<1$, a larger $f_0$ results in a smaller $S(T_N)$, which may cancel the effect of relocalization. Taking CePd$_2$Si$_2$ as an example, we predict $S(T_N)=0.62R\ln2$ using $T_N=9.9\,$K and $T^*=40\,$K \cite{Yang2008b}, quite close to the experimental value of $S(T_N)=0.68R\ln2$ \cite{Besnus1991,Sheikin2002}. This suggests a rough scaling relation between $S(T_o)$ and $T_o/T^*$ ($T_o=T_c$, $T_N$, $T_{L}$, etc), that can be easily applied to other materials in the future. 

\section{Conclusions}
In summary, our introduction of hybridization effectiveness into the two-fluid description yields a viable alternative to the Doniach phase diagram \cite{Doniach1977} for heavy electrons, while its determination from experiment leads to a unified phenomenological explanation of their magnetic properties. It will be important to extend the present analysis to a number of other heavy electron materials and to determine the expected magnetic field dependence of the hybridization effectiveness parameter, $f_0$. An analysis of future experiments should yield a hybridization-based three dimensional pressure/magnetic field phase diagram, tell us whether it is possible to place all heavy electron materials on it, and establish the localization line for those materials with $f_0>1$. Experiment may also provide further information on the physical origin of $f_0$ -- e.g., does it scale with $J$ and so mirror the Kondo coupling? At this stage the new improved two-fluid description presented here shows considerable promise as a candidate phenomenological standard model of their emergent behaviors and should provide the basis for the development of a microscopic theory that explains these as a natural consequence of the coupling of the local moment spin liquid to the conduction electron sea in which it is immersed. \\ \\

{\bf \large \noindent Acknowledgments}\\

Y.Y. is supported by NSF-China (Grant No. 11174339) and Chinese Academy of Sciences; D.P. has been supported by branch member contributions to ICAM and thanks colleagues at the Aspen Center for Physics for useful discussions during the writing of this paper.

\end{document}